\documentclass[prb,12pt]{revtex4}

\usepackage[english]{babel}
\usepackage[intlimits]{amsmath}
\usepackage[dvips]{graphicx}

\begin{document}

\begin{titlepage}

\title{Understanding Mechanochemical Coupling in Kinesins Using First-Passage Time Processes}
\author{Anatoly B. Kolomeisky, Evgeny B. Stukalin}
\affiliation{Department of Chemistry, Rice University, Houston, 
TX 77005-1892, USA}
\author{Alex A. Popov}
\affiliation{Department of Chemistry, Moscow State University, Moscow, Russia 117899}

\begin{abstract} 
Kinesins are processive motor proteins that move along microtubules in a stepwise manner, and their motion is powered by the hydrolysis of ATP. Recent experiments have investigated the  coupling between  the individual steps of single kinesin molecules and ATP hydrolysis, taking explicitly into account forward steps, backward steps and detachments. A theoretical study of mechanochemical coupling in kinesins, which extends the  approach used successfully to describe the dynamics of conventional motor proteins, is presented.  The possibility of irreversible detachments of kinesins from the microtubules is also explicitly taken into account. Using the method of first- passage times, experimental data on the mechanochemical coupling in kinesins are fully described using the simplest two-state model. It is shown that the dwell times for the kinesin to move one step forward or backward, or to dissociate irreversibly  are the same, although the probabilities of these events  are different. It is concluded that the current theoretical view, that only the forward motion of the motor protein molecule is coupled to ATP hydrolysis, is consistent  with all available experimental observations for kinesins.  

\end{abstract}

\maketitle


Running title: \hspace{1em} Understanding Mechanochemical Coupling in Kinesins.\\

\end{titlepage}

\section*{INTRODUCTION}

There are several classes of enzymes, called molecular motor proteins, that are critical for many biological processes, but especially they are important for cellular transport and motility, cell division,  and transfer of genetic information (Lodish et al., 1995; Bray, 2001; Howard, 2001). The motor proteins, such as kinesins, myosins, DNA and RNA polymerases, move in a stepwise motion   along rigid molecular tracks (microtubules, actin filaments, and DNA molecules). The motion of motor proteins is fueled by the hydrolysis of ATP or related compounds. However, the exact mechanism of the coupling between the chemical energy of hydrolysis and the mechanical motion of motor proteins is still unknown, and it remains one of the most important problems in biology.

Kinesins provide the most convenient system to investigate the mechanochemical coupling in motor proteins since biophysical, chemical and mechanical properties of these molecules are now well studied at single-molecule level (Howard et al., 1989; Svoboda et al., 1994; Schnitzer and Block, 1997; Kojima et al., 1997; Visscher et al., 1999; Schnitzer et al., 2000; Nishiyama et al., 2003; Asbury et al., 2003, Yildiz et al., 2003). Conventional kinesins are dimeric two-headed molecules, which hydrolyze ATP and move stochastically in 8.2-nm steps along the microtubules. These motor proteins can make hundreds of steps before dissociating from the microtubules and they can be processive  even against the opposing load as high as 7-8 pN (Visscher et al., 1999; Schnitzer et al., 2000; Nishiyama et al., 2002). Kinesin moves preferentially in the forward direction (plus end of microtubules),  however, at high loads the frequency of backward steps (in the direction of minus end of the microtubule) is increasing (Nishiyama et al., 2002). 

In order to understand how the motor proteins function, it is important to investigate how the chemical energy of ATP hydrolysis is transformed into the mechanical motion of proteins. To approach this fundamental problem, first, several critical questions should be answered: 1) How many ATP molecules consumed for each kinesin's step? 2) Are ATP molecules hydrolyzed for any step, forward or backward? 3) Is there a futile hydrolysis in kinesin motion, i.e., ATP consumption without actual moving of the motor protein?   

In recent experiments (Nishiyama et al., 2002), the mechanism of mechanochemical coupling in motor proteins has been studied by correlating the forward and backward movements of single kinesin molecules to the hydrolysis of ATP. Using optical trapping nanometry system, the time trajectories of single kinesin molecules have been measured for different external forces and for different ATP concentrations. It was found that the dwell times before the forward and backward steps are the same at all external forces and at all ATP concentrations. A biased Brownian motion model with asymmetric potentials was developed to explain the bidirectional motions of kinesins. Based on this model, it was concluded that the hydrolysis of single ATP molecule is coupled to either forward or backward steps of kinesins. 

Although the theoretical picture  presented by Nishiyama et al., 2002, that both forward and backward steps of kinesins are created by the same mechanochemical transduction mechanism, seems to be able to describe several features of the kinesin motility, there are  serious fundamental problems with this view. It contradicts  the current biochemical view of this process and earlier studies (Schnitzer and Block, 1997; Hua et al., 1997; Coy et al., 1999) that show a tight coupling, i.e., one ATP molecule is hydrolyzed per each forward 8-nm step. Note, however, that these earlier investigations mainly neglected the backward steps in their statistical analysis. In addition, the asymmetric potential used in the biased Brownian motion model breaks the periodic symmetry of the system, and it violates the principle of microscopic reversibility since the backward processes are not taken into account. Furthermore, this model cannot predict analytically the fraction of the forward and backward steps separately, and it also fails to account for irreversible detachments of kinesin molecules from the microtubules, which are observed in experiments. Clearly, a better  quantitative theoretical description, which does not violate the basic physical and chemical principles, is needed in order to satisfactorily understand the mechanochemical coupling in kinesins. The aim of this article is to discuss in detail such a theoretical approach.

We present a theoretical analysis of mechanochemical coupling and dynamics of kinesin molecules which utilizes  the first-passage time processes (van Kampen, 1997) in  periodic discrete-state stochastic models. This is an extension of the recently developed approach (Qian, 1997; Kolomeisky and Widom, 1998; Fisher and Kolomeisky, 1999a; Fisher and Kolomeisky, 1999b; Kolomeisky and Fisher, 2000a; Kolomeisky and Fisher, 2000b; Kolomeisky, 2001), which has been used successfully to analyze in detail the dynamics of single conventional kinesin molecules (Fisher and Kolomeisky, 2001) and myosin-V (Kolomeisky and Fisher, 2003). We argue that the experimental observations by  Nishiyama et al., 2002, can be described by the simplest ($N=2$)-state model with irreversible detachments, in which  ATP hydrolysis is tightly coupled only to the forward steps of motor proteins.

\section*{Theoretical Approach}

\subsection*{Chemical Kinetic Models}

Our approach is based on using multi-state discrete stochastic, or chemical kinetic, models. The main assumption of the simplest periodic sequential chemical kinetic model, which is shown in Fig. 1a, is that a motor protein molecule is viewed as a particle that moves along a periodic linear track from one binding site to the next one through the sequence of $N$ biochemical conformations. The particle in  state $j$ can jump forward to state $j+1$ with the rate $u_{j}$, or it can slide one step backward to the site $j-1$ with the rate $w_{j}$. After moving $N$ sites forward the motor protein comes to the same biochemical state but shifted by a step size distance $d$. For kinesins this distance is 8.2 nm, and it is equal to the size of a tubulin subunit in microtubules (Howard, 2001). The states $j=lN$ ($l=0, \pm 1, \pm2, \cdots$) represent the biochemical conformations where the motor protein molecule is tightly bound to the track, i.e., to the microtubule in case of kinesins, and without the ATP fuel molecule. ATP binding corresponds to the transitions from states $j=lN$ to $j=1+lN$, while other forward transitions describe the ATP hydrolysis and subsequent release of hydrolysis products. It is important to note that, although the motor protein moves preferentially in one direction, the reverse transitions cannot be ignored in any reasonable model of motor protein motility, and the backward steps are frequently observed experimentally (Schnitzer and Block, 1997;  Nishiyama et al., 2002).

In the periodic sequential multi-state stochastic model the dynamics of the motor protein can be viewed as the motion of the particle on a  periodic  one-dimensional lattice (with a period $N$). This observation allows one to derive an explicit  analytical expressions for the mean velocity $V(\{u_{j},w_{j}\})$,     
\begin{equation}
V=\lim_{t \rightarrow \infty} \frac{d \langle x(t) \rangle }{dt},
\end{equation}
in terms of transition rates  $\{u_{j},w_{j}\}$ for any value of $N$ (Fisher and Kolomeisky, 1999a; Fisher and Kolomeisky, 1999b). Here, $x(t)$ measures the position of the single molecule on the linear track. Specifically, the mean velocity is given by (Kolomeisky and Fisher, 2000a)
\begin{equation}
V=d \ \frac{1-\prod_{j=0}^{N-1}(w_{j}/u_{j})}{R_{N}}=d(u_{eff}-w_{eff}),
\end{equation}
where the effective  forward and backward rates are defined as
\begin{equation}
u_{eff}=1/R_{N}, \quad w_{eff}=\frac{\prod_{j=0}^{N-1} (w_{j}/u_{j})}{R_{N}},
\end{equation}
with 
\begin{equation}
R_{N}=\sum_{j=0}^{N-1} r_{j}, \quad r_{j}=\frac{1}{u_{j}} (1+\sum_{k=1}^{N-1} \prod_{i=j+1}^{j+k} w_{i}/u_{i}).
\end{equation}
Note also the periodicity of transition rates, i.e., $u_{j \pm N} =u_{j}$ and $w_{j \pm N} =w_{j}$.

Similar arguments can also be applied  to obtain  closed-form exact analytic formulae for the dispersion $D(\{u_{j},w_{j}\})$ (or effective diffusion constant) of the motion, which is defined as follows,
\begin{equation}
D=\frac{1}{2} \lim_{t \rightarrow \infty} \frac{d}{dt} [ \langle x^{2}(t) \rangle -  \langle x(t) \rangle^{2} ]. 
\end{equation}
The simultaneous knowledge of both the velocity $V$ and the dispersion $D$ determines the bounds on rate-limiting biochemical transitions and thus provides a valuable information about the mechanism of motor proteins motility (Visscher et al., 1999; Kolomeisky and Fisher, 2000a; Fisher and Kolomeisky, 2001).

One of the advantages of using chemical kinetic models to describe the processivity of motor proteins is the ability to easily incorporate the effect of external force $F$ (Fisher and Kolomeisky, 1999, 2001). This can be done with the introduction of load-distribution factors, $\theta_{j}^{+}$ and  $\theta_{j}^{-}$ ( for $j=0,1, \cdots, N-1$), that modify the transition rates in the following way,
\begin{eqnarray}
u_{j} & \Rightarrow & u_{j}(F)=u_{j}^{0} \exp(-\theta_{j}^{+}Fd/k_{B}T), \nonumber \\
w_{j} & \Rightarrow & w_{j}(F)=w_{j}^{0} \exp(\theta_{j}^{-}Fd/k_{B}T).
\end{eqnarray}
This is the consequence of the fact that the external load $F$ modifies the activation barriers for forward and backward transitions, and the load-distribution factors reflect how they changed. It is also  reasonable to assume that
\begin{equation}
\sum_{j=0}^{N-1} (\theta_{j}^{+}+\theta_{j}^{-})=1,
\end{equation}
since the motor protein, making a step $d$ against an external force $F$ and  going through $N$ intermediate steps, produces a work equal to $F d$. A force at which the motor protein stops moving is called a stall force.

\subsection*{First-Passage Time Processes}

In many single-molecule experiments on motor proteins the fractions of forward and backward steps and dwell times between the consecutive events are measured (Nishiyama et al., 2002; Asbury et al., 2003; Mehta et al. 1999). In terms of chemical kinetic models discussed above, these experimental quantities can be associated with the so-called splitting probabilities and conditional mean first-passage times, correspondingly. First-passage  processes for  sequential multi-state stochastic models are well studied (van Kampen, 1997; Pury and Caceres, 2003), and thus the available results can be easily adopted for the description of motor proteins dynamics. 

Consider a motor protein particle in state $j$, as shown in Fig. 1a. Recall that the sites $-N$, 0 and $N$ correspond to the binding sites for motor proteins. Now let us define $\pi_{j,N}(\{u_{j},w_{j}\})$ as the probability that the particle starting from state $j$ will reach the site $N$, before backtracking to the previous binding site $-N$. Similarly, we can define $\pi_{j,-N}(\{u_{j},w_{j}\})$ as the probability for the particle to advance to  state $-N$ for the first time before  reaching the forward binding site $N$. These quantities are called the splitting probabilities (van Kampen, 1997). We are mainly interested in the case of $j=0$, since the probabilities  $\pi_{0,\pm N}$ give us the forward and backward fractions of stepping for the motor protein particle. The explicit expressions for splitting probabilities are known (van Kampen, 1997), and for the periodic $N$-state stochastic models we obtain a simple relation,
\begin{equation}
\pi_{0,N}=1-\pi_{0,-N}=\frac{1}{1+\prod_{j=0}^{N-1} (w_{j}/u_{j})}.
\end{equation}

In a similar fashion, we can define the conditional mean first-passage times $\tau_{j,\pm N}(\{u_{j},w_{j}\})$, that represent the average time the  particle spends before advancing forward or backward to  sites $\pm N$, correspondingly.  Then it is  easy to conclude that the dwell times for the forward steps of motor proteins correspond to $\tau_{0,N}$, while the dwell times for the backward steps are given by  $\tau_{0,-N}$. The explicit expressions for  the dwell times within the periodic $N$-state chemical kinetic model can be  derived from more general equations that are not restricted by periodicity conditions (see Pury and Caceres, 2003), yielding
\begin{equation}
\tau_{0,N}=\frac{\pi_{0,N}}{u_{eff}}, \quad \tau_{0,-N}=\frac{\pi_{0,-N}}{w_{eff}},
\end{equation}
where the effective transition rates  $u_{eff}$ and $w_{eff}$ are defined in Eqs. (3) and (4).  Then applying  the Eqs. (3) and taking into account the relations for the forward and backward fractions [see Eq. (8)], we conclude that 
\begin{equation}
\tau_{0,N}=\tau_{0,-N}.
\end{equation}
This is a very important result because it indicates that the dwell times for the forward and backward steps are {\it always} equal to each other for {\it any} set of transition rates, although the probabilities of these steps may differ significantly. It is also important to note that  periodic conditions in the system  are crucial for this conclusion. 

\subsection*{Effect of Detachments}

Motor proteins do not always stay binded to the linear track, they can dissociate and diffuse away. For kinesins moving along the microtubules the effectively irreversible detachments have been observed  experimentally (Schnitzer et al., 2000; Nishiyama et al., 2002). Theoretically, the effect of detachments on the drift velocity, dispersion and the stall force has been investigated (Kolomeisky and Fisher, 2000a) using  the extension  of the simplest sequential multi-state stochastic   model.  However, to best of our knowledge, the problem of how the motor protein dissociations change the mean first-passage time processes, namely, the  fractions and  mean dwell times of forward and backward steps, has not been studied at all. Below we outline how this effect  can be solved by mapping it into another sequential multi-state stochastic model but {\it without} detachments, for which the results are already known.

Consider a motor protein particle in state $j$ as shown in Fig. 1b. It can move forward (backward) with the rate $u_{j}$ ($w_{j}$), or it can dissociate irreversibly with the rate $\delta_{j}$. We again define $\pi_{j,N}$ and $\pi_{j,-N}$ as splitting probabilities of reaching for the first time the forward (at $N$) or the backward (at $-N$) binding site. In addition, we introduce a new function $\pi_{j,\delta}$ as a probability for the motor protein, which starts at the site $j$, to detach before reaching the forward or the backward binding states. These probabilities are  related through the normalization condition,
\begin{equation}
\pi_{j,N}+\pi_{j,-N}+\pi_{j,\delta}=1.
\end{equation}

Now we may recall that the particle at the site $j$ has to jump to the site $j+1$  or $j-1$, or it will detach. These jumps have the probabilities $u_{j}/(u_{j}+w_{j}+\delta_{j})$,  $w_{j}/(u_{j}+w_{j}+\delta_{j})$ and $\delta_{j}/(u_{j}+w_{j}+\delta_{j})$, correspondingly. Then the expression for the forward splitting probability is given by (van Kampen, 1997)
\begin{equation}
\pi_{j,N}=\frac{u_{j}}{(u_{j}+w_{j}+\delta_{j})}\pi_{j+1,N}+\frac{w_{j}}{(u_{j}+w_{j}+\delta_{j})}\pi_{j-1,N},
\end{equation}
for any $-N<j<N$, and with the obvious choice of boundary conditions,
\begin{equation}
\pi_{N,N}=1, \quad \pi_{N,-N}=0.
\end{equation}
Similar equations can be derived for the backward splitting probabilities $\pi_{j,-N}$. 

The Eq. (12) can be easily rewritten as a difference equation, i.e.,
\begin{equation}
u_{j} \pi_{j+1,N}+w_{j} \pi_{j-1,N} -(u_{j}+w_{j}+\delta_{j}) \pi_{j,N}=0.
\end{equation}
Assume that the solution of this equation can be presented as
\begin{equation}
\pi_{j,N}= \phi_{j} \pi_{j,N}^{*},
\end{equation}
where the function $\pi_{j,N}^{*}$ is the splitting forward probability for a new system without detachments, and the auxiliary function $\phi_{j}$ is yet to be determined. Substituting Eq. (15) into the Eq. (14) we obtain,
\begin{equation}
u_{j} \phi_{j+1}\pi_{j+1,N}^{*}+w_{j} \phi_{j-1}\pi_{j-1,N}^{*} -(u_{j}+w_{j}+\delta_{j}) \phi_{j} \pi_{j,N}^{*}=0.
\end{equation}
If we define  new rates  for the stepping process without detachments as
\begin{equation}
u_{j}^{*} = u_{j} \phi_{j+1}, \quad w_{j}^{*}=w_{j} \phi_{j-1},
\end{equation}
and also require that
\begin{equation}
u_{j}^{*} +  w_{j}^{*}= u_{j} \phi_{j+1} + w_{j} \phi_{j-1}=(u_{j}+w_{j}+\delta_{j}) \phi_{j},
\end{equation}
then the Eq. (16) is easily transformed into
\begin{equation}
u_{j}^{*} \pi_{j+1,N}^{*}+w_{j}^{*} \pi_{j-1,N}^{*} -(u_{j}^{*}+w_{j}^{*}) \pi_{j,N}^{*}=0,
\end{equation}
with the boundary conditions $\pi_{N,N}^{*}=1$ and  $\pi_{-N,N}^{*}=0$. These boundary conditions  also mean that $\phi_{-N}=\phi_{N}=1$. Examining  Eq. (19), one can observe that this is the expression to determine the forward splitting probability of the  sequential multi-state stochastic process with rates $\{u_{j}^{*}, w_{j}^{*}\}$ but without detachments, for which the solutions are available (van Kampen, 1997). It leads to the explicit equation for the forward splitting probability. Similar arguments can be developed for the backward splitting probabilities.

Our analysis relies on the ability to compute the functions $\phi_{j}$, which can be accomplished by utilizing the Eq. (18). However, it is more convenient to look at $\phi_{j}$ as elements of the left eigenvector of a $(2N+1)$$\times$$(2N+1)$ matrix ${\mathbf M}$, for which the non-zero elements are given by
\begin{equation}
M_{ij}=\left \{\begin{array} {cc}
                -(u_{j}+w_{j}+\delta_{j}), & \mbox{for } i=j;   \\
                  w_{j},                   & \mbox{for } i=j-1; \\
                  u_{j},                   & \mbox{for } i=j+1;
                \end{array} \right.
\end{equation}
with $-N<i,j<N$.

The effect of detachments for conditional mean first-passage times can be investigated in a similar way. Here we again define $\tau_{j,N}$ ($\tau_{j,-N}$) as the mean  time to reach the forward (backward) binding state $N$ ($-N$) for the first time. In addition, we define $\tau_{j,\delta}$ as a mean first-passage time for the motor protein particle  to dissociate from the molecular track  before reaching the forward or backward binding sites $\pm N$. The mean first-passage times can be found by solving the backward Master Equation (see Pury and Caceres, 2003),
\begin{equation}
u_{j} \tau_{j+1,N}+w_{j} \tau_{j-1,N} -(u_{j}+w_{j}+\delta_{j,N}) \tau_{j,N}=-1,
\end{equation}
with the boundary conditions $\tau_{\pm N,N}=0$. Again, looking for the solution in the form $\tau_{j,N}=\phi_{j} \tau_{j,N}^{*}$, and using Eqs. (17) and (18), we obtain the following expression,
\begin{equation}
u_{j}^{*} \tau_{j+1,N}^{*}+w_{j}^{*} \tau_{j-1,N}^{*} -(u_{j}^{*}+w_{j}^{*}) \tau_{j,N}^{*}=-1,
\end{equation}
that determines the forward mean first-passage time for the system without detachments. Because the exact solutions for this case are available (van Kampen, 1997; Pury and Caceres, 2003) the expressions for the mean first-passage times for the system with detachments can be easily obtained.  

The general equations for splitting probabilities and mean first-passage times are quite complex, and we present in the next subsection the expressions only for simple cases $N=1$ and $N=2$. However, it can be shown that for  any $N$  the calculations of the mean dwell times to move forward,  backward or to dissociate leads to the following important relation,
\begin{equation}
\tau_{0,N}=\tau_{0,-N}=\tau_{0,\delta}.
\end{equation}
This is one of the main results of our theoretical analysis.

\subsection*{Results for $N=1$ and $N=2$ models}

To illustrate our method, let us consider two simple cases, $N=1$ and $N=2$ periodic sequential stochastic models with detachments. When the period of the system is $N=1$, the auxiliary function $\phi_{0}$ can be easily calculated,
\begin{equation}
\phi_{0}=\frac{u+w}{u+w+\delta},
\end{equation}
and also recall that $\phi_{-1}=\phi_{1}=1$. This leads to the simple relations for the splitting probabilities,
\begin{equation}
\pi_{0,1}=u/(u+w+\delta), \quad \pi_{0,-1}=w/(u+w+\delta), \quad \pi_{0,\delta}=\delta/(u+w+\delta);
\end{equation}
and for the mean first-passage times,
\begin{equation}
\tau_{0,1}=\tau_{0,-1}=\tau_{0,\delta}=1/(u+w+\delta).
\end{equation}

For $N=2$ case, the calculations become more tedious. The results for the functions $\phi_{-1}$, $\phi_{0}$, and $\phi_{1}$ are given by
\begin{equation}
\phi_{-1}=\frac{u_{0}u_{1}^{2}-u_{0}w_{1}^{2} +w_{1}(u_{0}+w_{0}+\delta_{0})(u_{1}+w_{1}+\delta_{1})}{ \left[(u_{0}+w_{0}+\delta_{0})(u_{1}+w_{1}+\delta_{1})-(u_{0}w_{1}+u_{1}w_{0}) \right](u_{1}+w_{1}+\delta_{1})};
\end{equation}
\begin{equation}
\phi_{0}=\frac{u_{0}u_{1}+w_{0}w_{1}}{ \left[(u_{0}+w_{0}+\delta_{0})(u_{1}+w_{1}+\delta_{1})-(u_{0}w_{1}+u_{1}w_{0}) \right]};
\end{equation}
\begin{equation}
\phi_{1}=\frac{w_{0}w_{1}^{2}-w_{0}u_{1}^{2} +u_{1}(u_{0}+w_{0}+\delta_{0})(u_{1}+w_{1}+\delta_{1})}{ \left[(u_{0}+w_{0}+\delta_{0})(u_{1}+w_{1}+\delta_{1})-(u_{0}w_{1}+u_{1}w_{0}) \right](u_{1}+w_{1}+\delta_{1})}.
\end{equation}
Then, after lengthy but straightforward calculations,  it can be shown that the splitting probabilities are
\begin{equation}
\pi_{0,2}=\frac{ u_{0} u_{1}} { \left[ u_{0}u_{1} + w_{0}w_{1} + \delta_{0} \delta_{1} + \delta_{0}(u_{1}+w_{1}) + \delta_{1}(u_{0}+w_{0}) \right]};
\end{equation}
\begin{equation}
\pi_{0,-2}=\frac{ w_{0} w_{1}} { \left[ u_{0}u_{1} + w_{0}w_{1} + \delta_{0} \delta_{1} + \delta_{0}(u_{1}+w_{1}) + \delta_{1}(u_{0}+w_{0}) \right]};
\end{equation}
and $\pi_{0,\delta}=1-\pi_{0,2}-\pi_{0,-2}$. Similar calculations for the mean first-passage times yield
\begin{equation}
\tau_{0,2}=\tau_{0,-2}=\tau_{0,\delta}=\frac{ u_{0} + u_{1} + w_{0} + w_{1} +\delta_{1}} { \left[ u_{0}u_{1} + w_{0}w_{1} + \delta_{0} \delta_{1} + \delta_{0}(u_{1}+w_{1}) + \delta_{1}(u_{0}+w_{0}) \right]}.
\end{equation}

\section*{Analysis of Kinesin Data}

Structural, biochemical and kinetic data on kinesins suggest that the protein molecule goes through at least four intermediate states (Lodish et al., 1995, Bray, 2001). However, recent study of kinesin dynamics using ($N=2$)-state chemical kinetic model, which takes into account the irreversible detachments, provides a very reasonable description of some aspects of mechanochemical coupling in this system (Fisher and Kolomeisky, 2001). Thus, in order to analyze the experimental data of Nishiyama et al., 2002, we adopt the simplest model which includes only two states. The states $j=\cdots,-2,0,2,\cdots$ would correspond to the kinesin with both molecular heads tightly bound to the microtubule and without ATP molecule. The states $j=\cdots,-1,1,\cdots$ label all other kinesin conformations after ATP binding and subsequent hydrolysis and release of its products.

It now follows that the forward ATP-binding transition depends linearly on ATP concentration, $u_{0}^{0}=k_{0}^{0}$[ATP], where the superscript 0 indicates the case of zero load: see also Eq. (6). At the same time the next forward rate $u_{1}$ and the backward rate $w_{1}$ do not depend on ATP concentration, while they may  change under the effect of external forces.

The final backward rate $w_{0}$ might, in principle, depend on concentrations of ADP and inorganic phosphate that both are the products of ATP hydrolysis. However, most current experiments on kinesins utilize ATP regeneration system (Svoboda et al., 1994; Visscher et al, 1999; Schnitzer et al., 2000; Nishiyama, 2002), in which there are no independent control of [ADP] and [P$_{i}$]. As a result, we adopt a phenomenological description of this backward transition, namely,  
\begin{equation}
w_{0}^{0}=k_{0}'\mbox{[ATP]}/(1+\mbox{[ATP]}/c_{0})^{1/2},
\end{equation}
where the parameter $c_{0}$ effectively describes the ATP regeneration process. This approach has been used successfully to describe the mechanochemical transitions in kinesin and myosin-V (Fisher and Kolomeisky, 2001; Kolomeisky and Fisher, 2003). Note, however, that the specific description of ATP regeneration process has a minimal effect in fitting of experimental results. 

After systematically exploring the multi-dimensional space of parameters and using Eqs. (30)-(33) the fractions of forward and backward steps and mean dwell times between the consecutive steps of kinesins can be well described by  the following rate constants
\begin{eqnarray}
k_{0}^{0} \simeq 5.1 \mu\mbox{M}^{-1}\mbox{s}^{-1}, \quad  k_{0}' \simeq 2.8 \mu\mbox{M}^{-1}\mbox{s}^{-1}, \quad  c_{0} \simeq 1.7 \mu\mbox{M}, & \  w_{1}^{0} \simeq 5.5\cdot 10^{-4} \mbox{s}^{-1},  \nonumber \\
u_{1}^{0} \simeq 121 \mbox{s}^{-1}, \quad \delta_{0}^{0} \simeq 1.1 \mbox{s}^{-1}, \quad \delta_{1}^{0} \simeq 1.6 \cdot 10^{-3} \mbox{s}^{-1}, & 
\end{eqnarray}
and load-distribution parameters
\begin{eqnarray}
 \theta_{0}^{+} \simeq 0.0, \quad  \theta_{0}^{-} \simeq 0.391, \quad \theta_{1}^{+} \simeq 0.086, &  \nonumber \\
 \theta_{1}^{-} \simeq 0.523, \quad \theta_{0}^{\delta} \simeq 0.047, \quad \theta_{1}^{\delta} \simeq 0.466. &
\end{eqnarray}
The results of the fitting of experimental observations are given in Figs. 2 and 3. Note that the values for the parameters reported here are in a good  agreement with the other independent investigation of kinesin motility (Fisher and Kolomeisky, 2001), where  the  multi-state periodic stochastic models have been used to analyze the single-molecule experimental measurements  of  velocities, stall forces and dispersions (Visscher et al., 1999).

\section*{Discussion}

Our theoretical analysis provides explicit expressions for the fractions of forward and backward steps and dissociations, and for the mean dwell times between consecutive steps of motor proteins. This allows us to investigate the problem of  mechanochemical coupling between the motion of kinesins  and ATP hydrolysis. Our main conclusion is that the mean dwell times to move forward, backward or irreversible detach are equal to each other independently of ATP concentration or external force. It means that the picture of  tight coupling  between ATP hydrolysis and forward steps of kinesins does not contradict the experimental findings of Nishiyama et al., 2002. Moreover, the proposed bidirectional biased model (Nishiyama et al., 2002), which assumes that a hydrolysis of a single ATP molecule is coupled to either forward or the backward movement, is basically incorrect since it violates the principle of microscopic reversibility and breaks the symmetry of the system. 

Our theoretical results could also be understood in the following way. The mean dwell times between  movements measured in single-molecule experiments are actually correspond to the mean lifetimes of states when the motor protein binds strongly to the linear track. Then these lifetimes should be independent of what direction the motor protein will go in the next step, although the probability of these steps might be rather different. 

The analysis of mean dwell times at different external forces, as shown in Fig. 3, suggests that there is a maximum at high loads. This maximum is close but not exactly at the stall force. When [ATP]=10 $\mu$M the maximum can be found at $F \simeq 6.6$ pN, while the stall force is approximately equal to 6.8 pN. At high ATP (1 mM) the position of maximum is shifted to 7.7 pN, with the calculated stall force $F_{S} \simeq 9.2$ pN. This can be understood in the following way. The external load decreases the forward transition rates, while slowing down the backward transitions. These two tendencies have an opposite effect on mean dwell times, and it leads to the observation of maximum at some specific value of external force.

Because our method provides exact expressions for biophysical parameters, we are able to study the effects of ATP concentration and external forces on these parameters, and we can make a qualitative predictions that can be checked experimentally.     

First, we investigate how mean dwell times depend on [ATP] at different external loads. As shown in Fig. 4, the larger the external force, the larger is the mean dwell time. However,  at constant force, the mean dwell time decreasing with increase in concentration of ATP. This is in agreement with intuitive expectations since at large [ATP] the binding process is faster. At the same time the external force slows down the binding and other forward processes more than it accelerates the backward transitions. These observations are also consistent with theoretical investigation of processivity of motor proteins using thermal ratchet approach (Parmeggiani et al., 2001). 

The  dependence of the fractions of different movements on ATP concentration at different external loads is presented in Fig. 5. The increase in [ATP] increases the probability of the forward steps, while making the fractions of backward steps and detachments negligible. Finally, the predictions for the force-velocity based on the fitted parameters are given in Fig. 6. These predictions are generally agree with  the values of drift velocities and stall forces obtained in other single-molecule experiments on kinesins (Visscher et al., 1999). However, the shapes for force-velocity curves are differ for ATP saturating conditions.

\section*{Conclusions}

In summary, we have presented a theoretical study of mechanochemical coupling in kinesins. The analysis of multi-state stochastic models of motility using the method of first-passage times allowed us to obtain the explicit formulae for fractions of steps in different directions,  and for the mean dwell times between the steps,  including the irreversible detachments. The experimental data on kinesins can be well described by this approach. Our analysis is consistent with the current theoretical view of tight coupling between catalytic cycles and mechanical steps for kinesins, i.e., one ATP molecule is hydrolyzed per each forward step, and the rare backward steps correspond to ATP production. Although our theoretical approach seems to provide a reasonable and convenient framework for investigating the mechanochemical coupling in different motor proteins, further experiments are needed in order to validate our theoretical picture.

\section*{Acknowledgments}

We acknowledge the support from the Camille and Henry Dreyfus New faculty Awards Program (under grant NF-00-056), the Welch Foundation (under grant C-1559), and US National Science Foundation through grant CHE-0237105. We also thank M.E. Fisher and Hong Qian  for critical discussions, suggestions  and encouragements.

\newpage

\hspace{5cm}{\bf FIGURE LEGENDS}
\vspace{1cm}

\noindent \mbox{FIGURE  1}. \hspace{0.4em} a) General schematic view of periodic multi-state stochastic models. A motor protein particle in state $j$ can make a forward transition at rate $u_{j}$, or it can undertake a backward transition at the rate $w_{j}$. The states $j=\cdots,-N,0,N,\cdots$ correspond to the strongly bound states. b) General scheme of periodic multi-state stochastic models with irreversible detachments. The particle in state $j$ can dissociate with a rate $\delta_{j}$.\\
\\
\noindent \mbox{FIGURE  2}. \hspace{0.4em} Probabilities, or fractions, of forward steps (circles), backward steps (triangles) and detachments (squares) as a function of the external force at a) [ATP]=1mM;  b) [ATP]=10$\mu$M. \\
\\
\noindent \mbox{FIGURE  3}. \hspace{0.4em} Dwell times between the adjacent movements of the kinesin molecule as a function of external force. The filled symbols correspond to experimental measurements at [ATP]=10$\mu$M, while open symbols describe the experiments at [ATP]=1mM. The circles mark the experimental measurements for dwell times before the forward steps; the triangles correspond to experimental dwell times before the backward steps; and squares describe the dwell times before detachments.\\
\\
\noindent \mbox{FIGURE  4}. \hspace{0.4em} Predictions for the dwell times as a function of [ATP]  at low ($F=1$ pN) and high external load ($F=5$ pN).\\
\\
\noindent \mbox{FIGURE  5}. \hspace{0.4em} Predictions for the variation of the fractions of forward steps, backward steps and detachments at a) $F=1$ pN; and b) $F=5$ pN.\\
\\
\noindent \mbox{FIGURE  6}.  \hspace{0.4em} Predictions for the force-velocity curves at different [ATP].\\
\\

\begin{figure}[t]
\begin{center}
\vskip 1.5in
\unitlength 1in
\begin{picture}(3.5,2.0)
\resizebox{3.25in}{1.6in}{\includegraphics{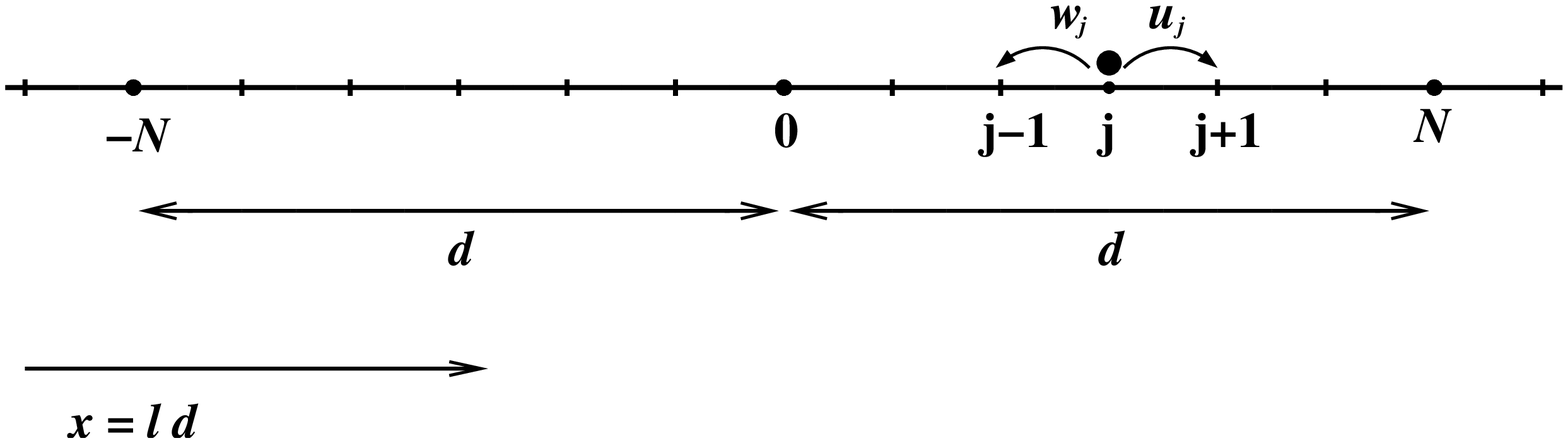}}
\end{picture}
\vskip 3in
 \begin{Large} Fig.1a   \end{Large}
\end{center}
\vskip 3in
\end{figure}

\begin{figure}[t]
\begin{center}
\vskip 1.5in
\unitlength 1in
\begin{picture}(3.5,2.0)
\resizebox{3.25in}{1.6in}{\includegraphics{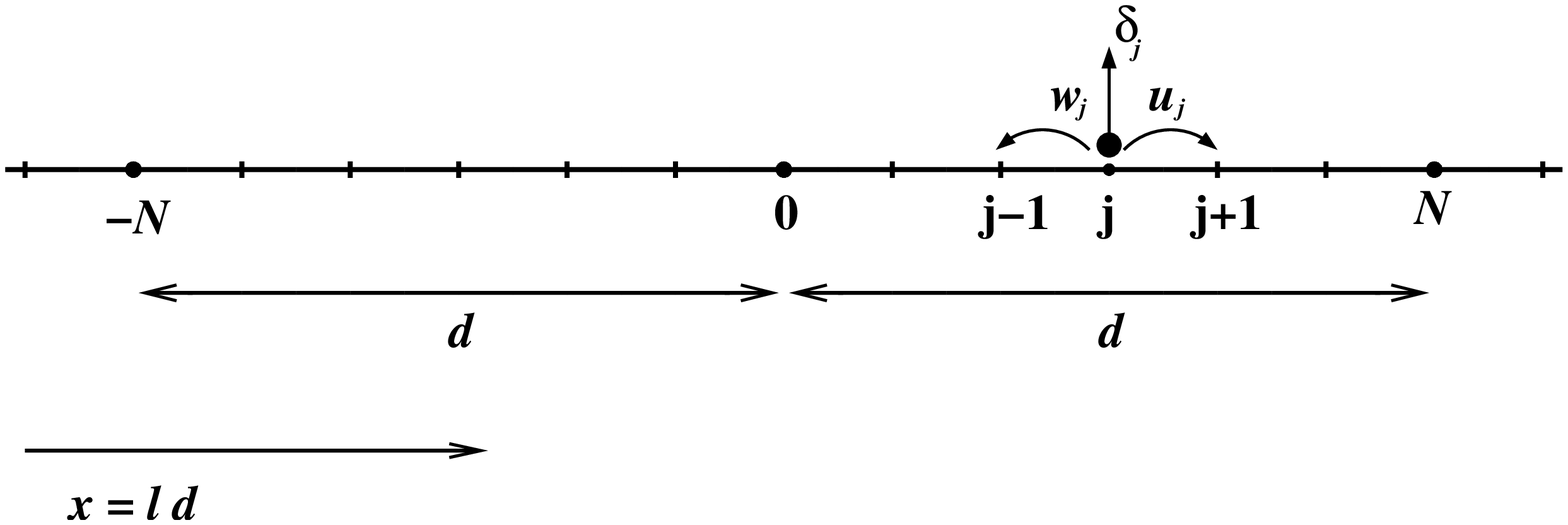}}
\end{picture}
\vskip 3in
 \begin{Large} Fig.1b   \end{Large}
\end{center}
\vskip 3in
\end{figure}

\begin{figure}[t]
\begin{center}
\vskip 1.5in
\unitlength 1in
\begin{picture}(3.5,2.0)
\resizebox{3.25in}{1.9in}{\includegraphics{Fig2a.eps}}
\end{picture}
\vskip 3in
 \begin{Large} Fig.2a   \end{Large}
\end{center}
\vskip 3in
\end{figure}

\begin{figure}[t]
\begin{center}
\vskip 1.5in
\unitlength 1in
\begin{picture}(3.5,2.0)
\resizebox{3.25in}{1.9in}{\includegraphics{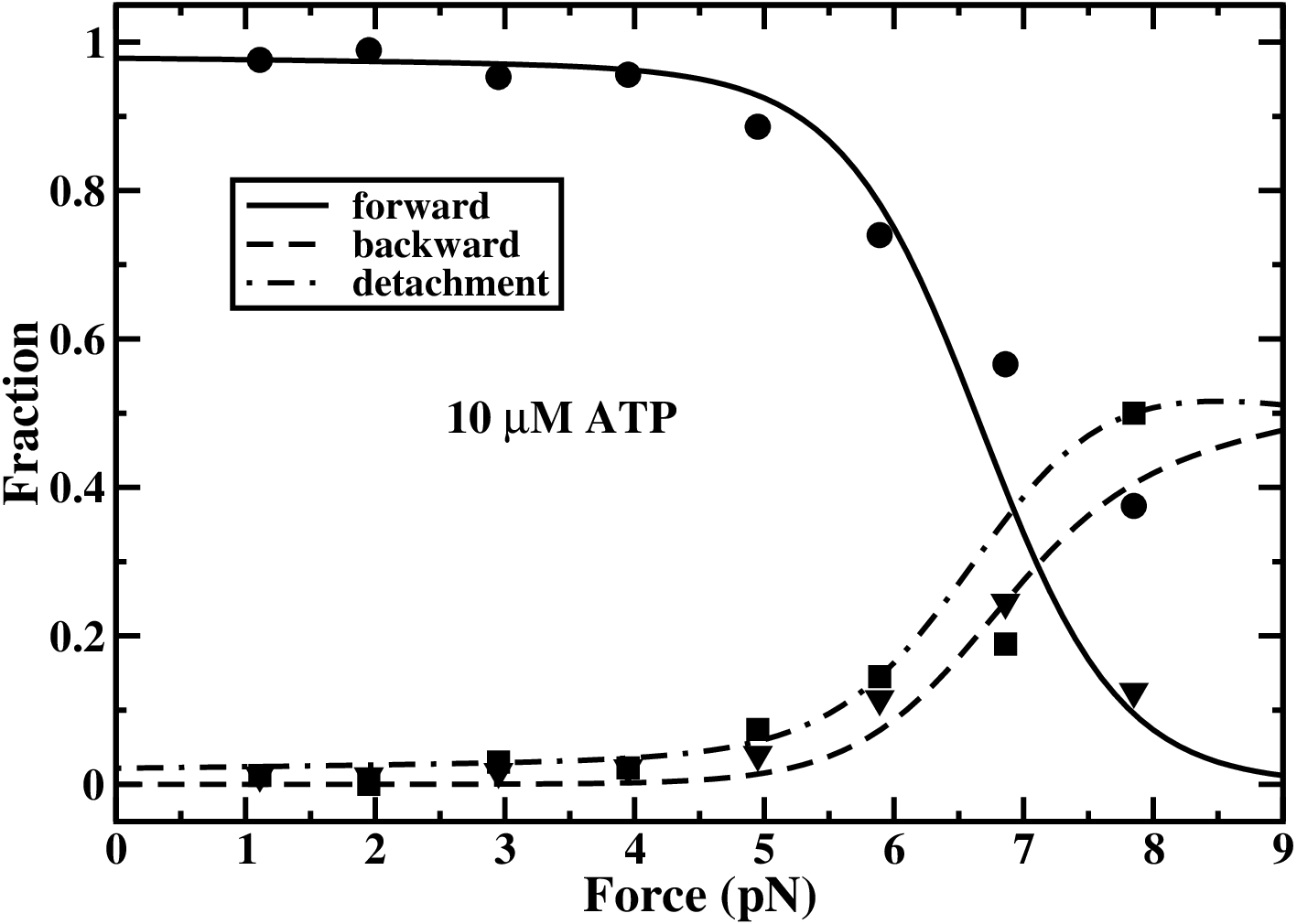}}
\end{picture}
\vskip 3in
 \begin{Large} Fig.2b   \end{Large}
\end{center}
\vskip 3in
\end{figure}

\begin{figure}[t]
\begin{center}
\vskip 1.5in
\unitlength 1in
\begin{picture}(3.5,2.0)
\resizebox{3.25in}{1.9in}{\includegraphics{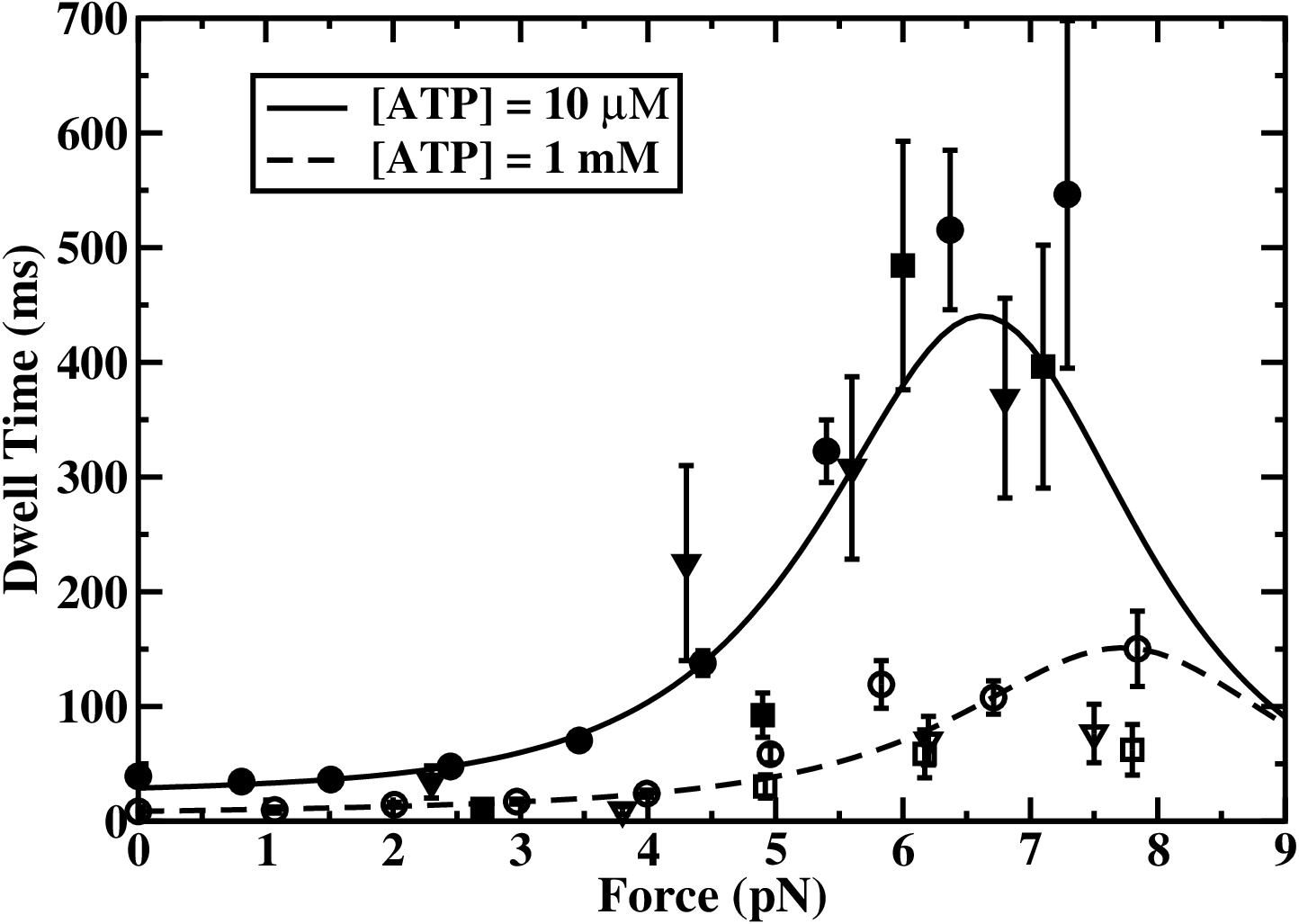}}
\end{picture}
\vskip 3in
 \begin{Large} Fig.3   \end{Large}
\end{center}
\vskip 3in
\end{figure}

\begin{figure}[t]
\begin{center}
\vskip 1.5in
\unitlength 1in
\begin{picture}(3.5,2.0)
\resizebox{3.25in}{1.9in}{\includegraphics{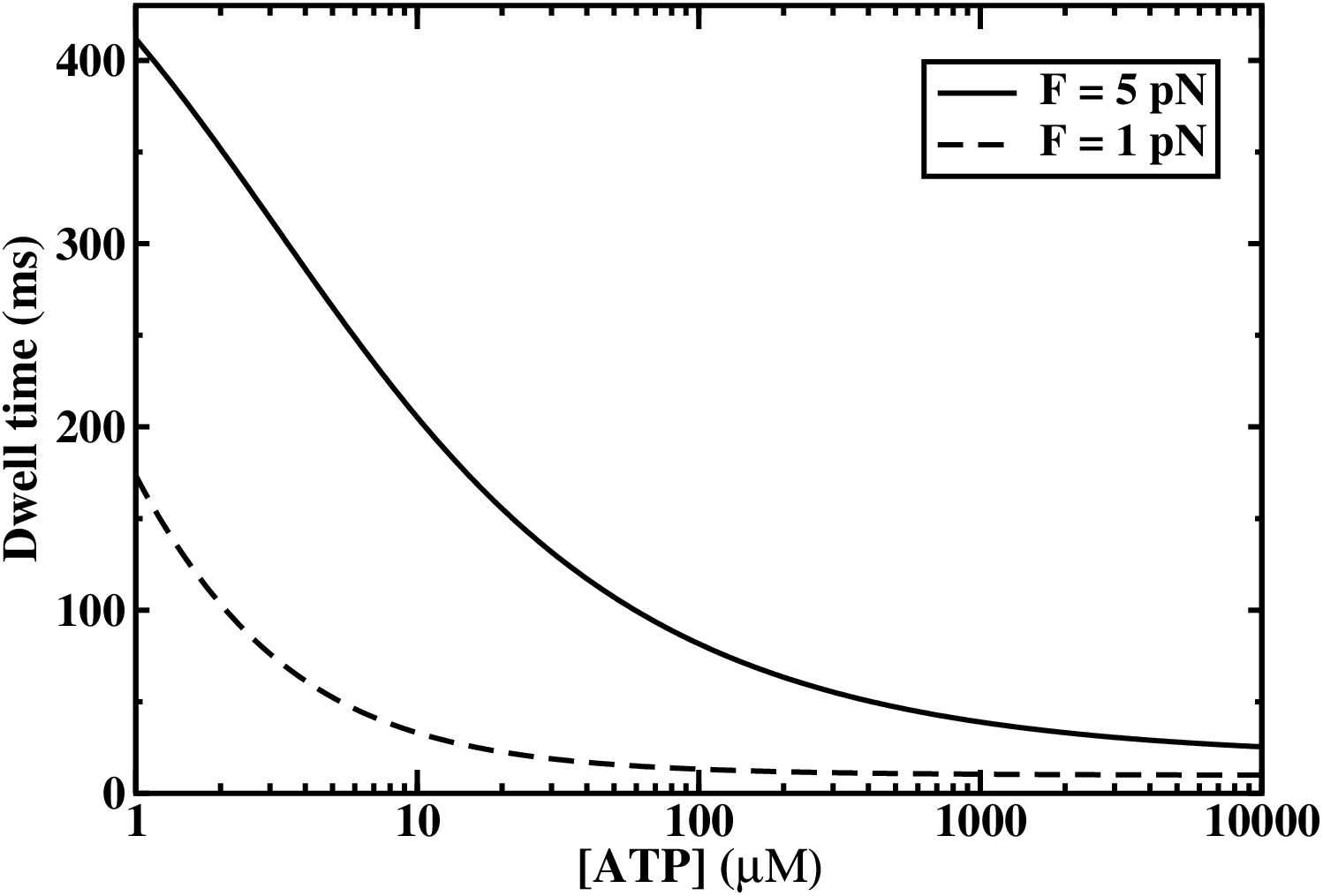}}
\end{picture}
\vskip 3in
 \begin{Large} Fig.4   \end{Large}
\end{center}
\vskip 3in
\end{figure}

\begin{figure}[t]
\begin{center}
\vskip 1.5in
\unitlength 1in
\begin{picture}(3.5,2.0)
\resizebox{3.25in}{1.9in}{\includegraphics{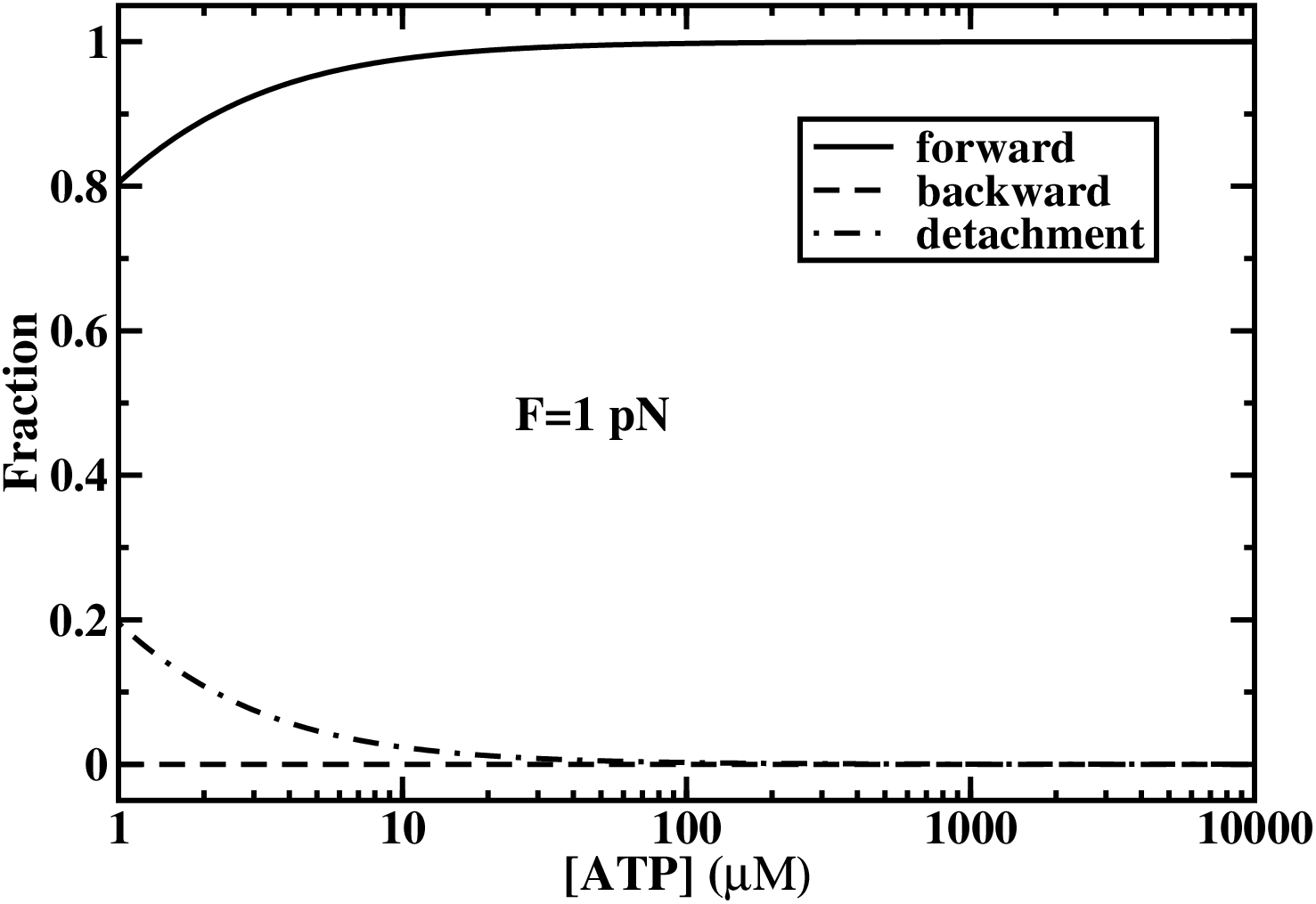}}
\end{picture}
\vskip 3in
 \begin{Large} Fig.5a   \end{Large}
\end{center}
\vskip 3in
\end{figure}

\begin{figure}[t]
\begin{center}
\vskip 1.5in
\unitlength 1in
\begin{picture}(3.5,2.0)
\resizebox{3.25in}{1.9in}{\includegraphics{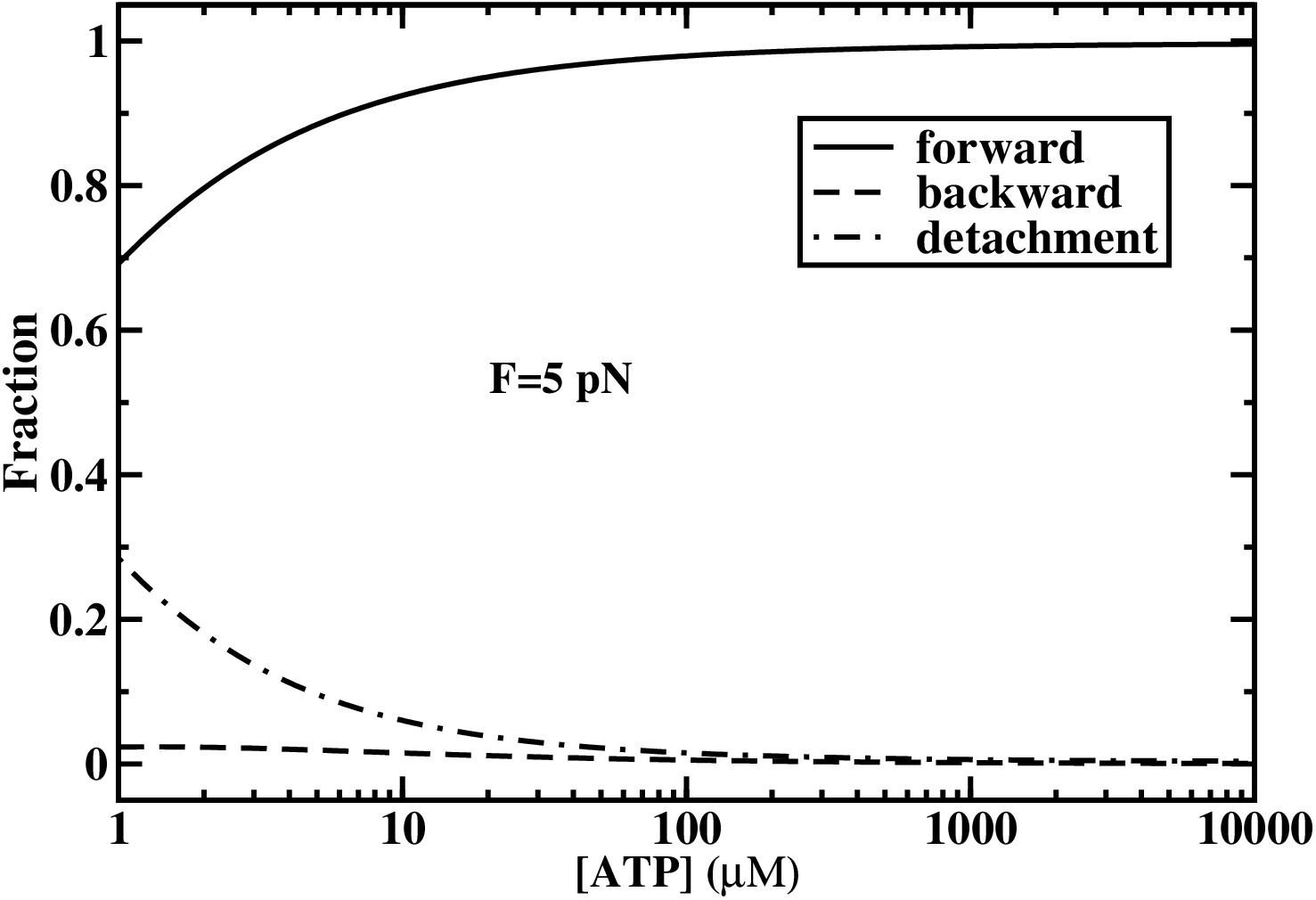}}
\end{picture}
\vskip 3in
 \begin{Large} Fig.5b   \end{Large}
\end{center}
\vskip 3in
\end{figure}

\begin{figure}[t]
\begin{center}
\vskip 1.5in
\unitlength 1in
\begin{picture}(3.5,2.0)
\resizebox{3.25in}{1.9in}{\includegraphics{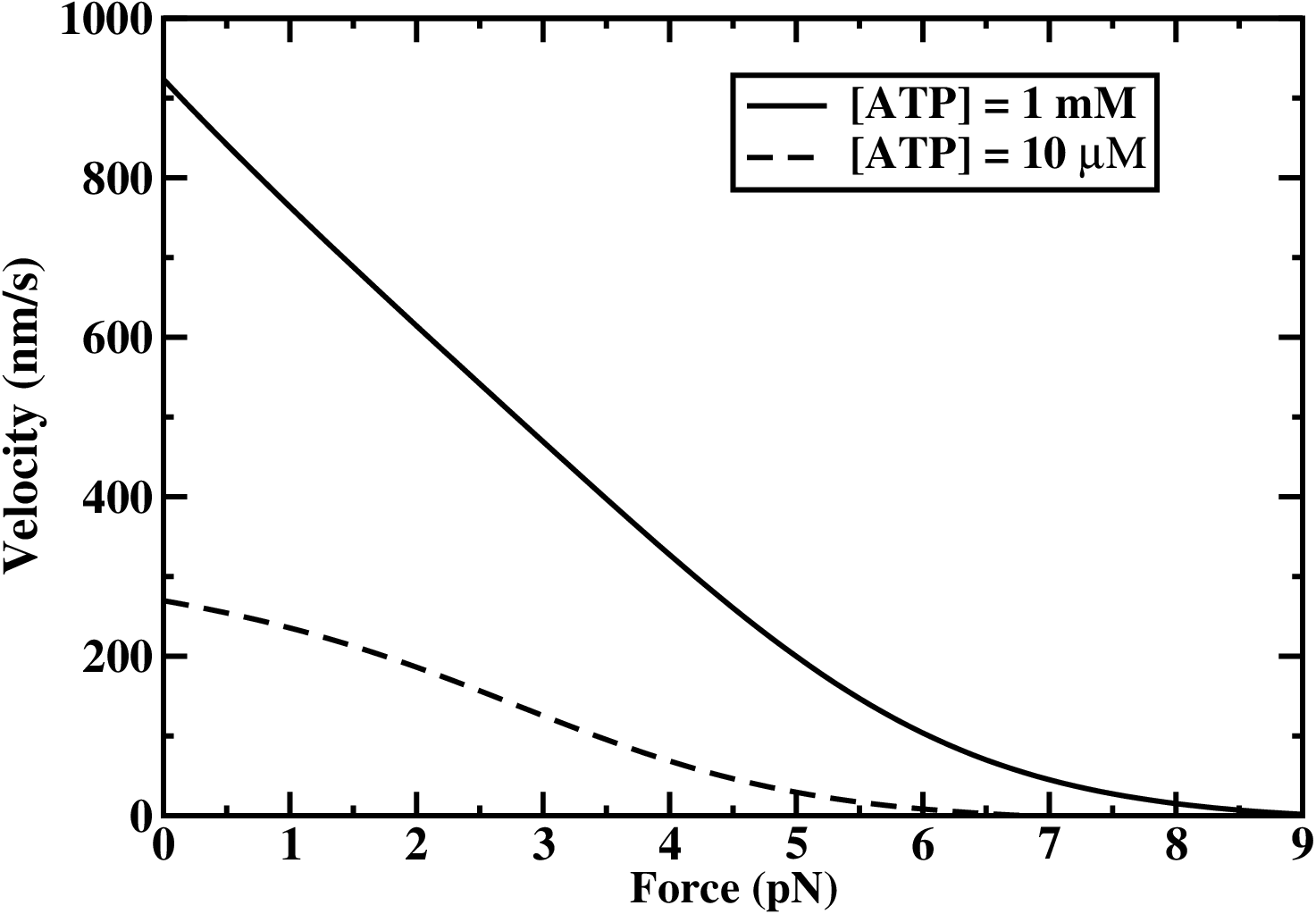}}
\end{picture}
\vskip 3in
 \begin{Large} Fig.6   \end{Large}
\end{center}
\vskip 3in
\end{figure}


\section*{References}

\begin{thebibliography}{99}


\bibitem[ ]{asbury} Asbury, C.L., Fehr, A.N., and S.M. Block. 2003. Kinesin Moves by an Asymmetric Hand-over-Hand Mechanism. {\it Science} 302: 2130-2134.
 
\bibitem[ ]{bray} Bray, D. 2001. Cell Movements: from molecules to motility, 2nd Edn. (Garland Publishing, New York) Chap. 5.

\bibitem[ ]{coy} Coy, D.L., M. Wagenbach, and J. Howard. 1999. Kinesin takes one 8-nm step for each ATP that it hydrolyzes. {\it J. Biol. Chem.} 274: 3667-3671.

\bibitem[ ]{FK} Fisher, M.E., and A.B. Kolomeisky. 1999(a). The force exerted by a molecular motor.  {\it Proc. Natl. Acad. Sci. USA} 96: 6597-6602. 

\bibitem[ ]{FK1} Fisher, M.E., and A.B. Kolomeisky. 1999(b). Molecular motors and the forces they exert. {\it Physica A} 274: 241-266. 

\bibitem[ ]{FK2} Fisher, M.E.,  and A.B. Kolomeisky. 2001. Simple mechanochemistry describes the dynamics of kinesin molecules.  {\it Proc. Natl. Acad. Sci. USA} 98: 7748-7753. 

\bibitem[ ]{howard1} Howard, J., A.J. Hudspeth, and R.D. Vale. 1989. Movements of microtubules by single kinesin molecules. {\it Nature} 342: 154-158.

\bibitem[ ]{howard} Howard, J. 2001. Mechanics of Motor Proteins and the Cytoskeleton. Sinauer Associates, Sunderland, Mass.

\bibitem[ ]{hua} Hua, W., E.C. Young, M.L. Fleming, and J. Gelles. 1997. Coupling of kinesin steps to ATP hydrolysis. {\it Nature} 388: 390-393.

\bibitem[ ]{kojima} Kojima, H.,  E. Muto, H. Higuchi, and T. Yanagida. 1997. Mechanics of single kinesin molecules measured by optical trapping nanometry. {\it Biophys. J.} 73: 2012-2022.

\bibitem[ ]{kolom} Kolomeisky, A.B. 2001. Exact results for parallel-chain kinetic models of biological transport. {\it J. Chem. Phys.} 115: 7253-7259.

\bibitem[ ]{KW} Kolomeisky, A.B., and B. Widom. 1998. A simplified ``ratchet model of molecular motors.'' {\it J. Stat. Phys.} 93: 633-645.

\bibitem[ ]{KF00a} Kolomeisky, A.B.,  and M.E. Fisher. 2000(a). Periodic sequential kinetic models with jumping branching and deaths. {\it Physica A} 279: 1-20.

\bibitem[ ]{KF00b} Kolomeisky, A.B., and M.E. Fisher. 2000(b). Extended kinetic models with waiting-time distributions: Exact results. {\it J. Chem. Phys.} 113: 10867-10877. 

\bibitem[ ]{KF03} Kolomeisky, A.B., and M.E. Fisher. 2003. A Simple kinetic model describes the processivity of myosin-V. {\it Biophys. J.} 84: 1642-1650.

\bibitem[ ]{lodish} Lodish, H., A. Berk, S.L. Zipursky, and P. Matsudaira. 1995. Molecular Cell Biology, 3rd ed. Scientific American Books, New York.

\bibitem[ ]{mehta} Mehta A.D., R.S. Rock, M. Rief, J.A. Spudich, M.S. Mooseker, and R.E. Cheney. 1999. Myosin-V is a processive actin-based motor. {\it Nature} 400: 590-593.

\bibitem[ ]{nishiyama} Nishiyama, M., H. Higuchi, and T. Yanagida. 2002. Chemomechanical coupling of the forward and backward steps of single kinesin molecules. {\it Nature Cell Bio.} 4: 790-797.

\bibitem[ ]{parmeggiani} Parmeggiani, A., F. J\"{u}licher, L. Peliti, and J. Prost. 2001. Detachment of molecular motors under tangential loading. {\it Europhys. Lett.} 56: 603-609.

\bibitem[ ]{pury} Pury, P.A., and M.O. Caceres. 2003. Mean first-passage and residence times of random walks on asymmetric disordered chains. {\it J. Phys. A: Math. Gen.} 36: 2695-2706.

\bibitem[ ]{qian} Qian, H. 1997. A simple theory of motor protein kinetics and energetics. {\it Biophys. Chem.} 67: 263-267.

\bibitem[ ]{schnitzer} Schnitzer, M.J., and S.M. Block. 1997. Kinesin hydrolyzes one ATP per 8-nm step. {\it Nature} 388: 386-390.

\bibitem[ ]{schnitzer1} Schnitzer, M.J., K. Visscher, and S.M. Block. 2000. Force production by single kinesin motors. {\it Nat. Cell. Biol.} 2: 718-723.


\bibitem[ ]{svoboda} Svoboda, K., P.P. Mitra, and S.M. Block. 1994. Fluctuation analysis of motor protein movement and single enzyme kinetics.  {\it Proc. Natl. Acad. Sci. USA} 91: 11782-11786.


\bibitem[ ]{vankampen} van Kampen, N.G., 1997. Stochastic Processes in Physics and Chemistry. 2nd Edn. Chap. 12, Elsevier. Amsterdam.


\bibitem[ ]{visscher} Visscher, K.,  M.J. Schnitzer, and S.M. Block. 1999. Single kinesin molecules studied with a molecular force clamp. {\it Nature} 400: 184-189.

\bibitem[ ]{yildiz} Yildiz, A., M. Tomishige, R.D. Vale, and P.R. Selvin. 2003. Kinesin Walks Hand-Over-Hand. {\it Science} 302: 676-678.

\end{thebibliography}
\end{document}